\documentclass{ieeeaccess}
\usepackage{cite}
\usepackage{amsmath,amssymb,amsfonts}
\usepackage{graphicx}
\usepackage[hidelinks]{hyperref}
\usepackage{multirow}
\usepackage{xcolor}
\usepackage{colortbl}
\usepackage{enumitem}
\usepackage{dblfloatfix}
\usepackage{ulem}
\usepackage{hhline}
\usepackage{float}
\usepackage{algpseudocode}
\usepackage{algorithm}
\usepackage{adjustbox}
\def\BibTeX{{\rm B\kern-.05em{\sc i\kern-.025em b}\kern-.08em
    T\kern-.1667em\lower.7ex\hbox{E}\kern-.125emX}}
\begin{document}
\history{Date of publication xxxx 00, 0000, date of current version xxxx 00, 0000.}
\doi{}

\title{Trading Off Scalability, Privacy, and Performance in Data Synthesis}
\author{
\uppercase{Xiao~Ling}\authorrefmark{1},
\uppercase{Tim~Menzies}\authorrefmark{2},
\uppercase{Christopher~Hazard}\authorrefmark{3},
\uppercase{Jack~Shu}\authorrefmark{4}, and
\uppercase{Jacob~Beel}\authorrefmark{5}
}

\address[1]{North Carolina State University, Raleigh, NC  27695 USA (e-mail: xling4@ncsu.edu)}
\address[2]{North Carolina State University, Raleigh, NC 27695 USA (e-mail: timm@ieee.org)}
\address[3]{Howso, Raleigh, NC 27603 USA}
\address[4]{Howso, Raleigh, NC 27603 USA}
\address[5]{Howso, Raleigh, NC 27603 USA}

\tfootnote{This work is a report of  an industrial collaboration between Howso and NCState. For full details, see ``Acknowledgments''.}

\markboth
{Author \headeretal: Preparation of Papers for IEEE TRANSACTIONS and JOURNALS}
{Author \headeretal: Preparation of Papers for IEEE TRANSACTIONS and JOURNALS}

\corresp{Corresponding author: Xiao~Ling (e-mail: xling4@ncsu.edu).}

\begin{abstract}
Synthetic data has been widely applied in the real world recently. One typical example is the creation of synthetic data for privacy concerned datasets. In this scenario, synthetic data substitute the real data which contains the privacy information, and is used to public testing for machine learning models. Another typical example is the unbalance data over-sampling which the synthetic data is generated in the region of minority samples to balance the positive and negative ratio when training the machine learning models. In this study, we concentrate on the first example, and introduce (a) the Howso engine, and (b) our proposed random projection based synthetic data generation framework. We evaluate these two algorithms on the aspects of privacy preservation and accuracy, and compare them to the two state-of-the-art synthetic data generation algorithms DataSynthesizer and Synthetic Data Vault. We show that the synthetic data generated by Howso engine has good privacy and accuracy, which results the best overall score. On the other hand, our proposed random projection based framework can generate synthetic data with highest accuracy score, and has the fastest scalability.
\end{abstract}


\titlepgskip=-21pt

\maketitle

\section{Introduction}\label{introduction}
\PARstart{T}{o} successfully apply artificial intelligent approaches (e.g. machine learning \& deep learning algorithms) in the real world application, data has became the most important part to support these algorithms. However, many types of data have privacy concerns and limited in terms of publicity~\cite{dahmen2019synsys, dankar2021fake}. This causes the issue on training the machine learning models on these regions since these models require support from huge training data. 

To mitigate this issue, synthetic data has been widely studied to substitute the real data. Synthetic data is the fake data points which generated by the generative model with the information from the real data points~\cite{smith2009improving, hernandez2022synthetic, assefa2020generating}. To evaluate the real world application of synthetic data, privacy preservation is one of the critical measurements which checks if identities from the original dataset can be detected or recognized in the synthetic data~\cite{liu2022privacy, rankin2020reliability, yale2020generation}. Similarity is another measurement since the synthetic data need to capture the information from the original data~\cite{james2021synthetic, nowok2015utility}. Moreover, another important measurement is to check if the synthetic data can be used to substitute the real data on training the models~\cite{fiestas2021rpa, hong2021synthetic}.

In this study, we  evaluate a new data synthesis method called  recursive random projections. Initially developed for optimization, recursive random projection  uses the FASTMAP~\cite{faloutsos1995fastmap, platt2005fastmap} technique to recursively bi-cluster the data into numerous small leaf  clusters. An optimizer then samples just $N=2$ points per leaf. But after reading the data synthesis literature~\cite{smith2009improving, hernandez2022synthetic, assefa2020generating, liu2022privacy, rankin2020reliability, yale2020generation,james2021synthetic, nowok2015utility,fiestas2021rpa, hong2021synthetic} we began to wonder if recursive random projection could also be a data synthesis algorithm just by sampling much more than $N=2$.

To check this, we perform the study described in this paper. 
As described in \S\ref{randomprojection}, 
random projection is augmented with mutation and crossover operators to generate synthetic data points per leaf cluster.
This is then compared to state-of-the-art synthetic data generation algorithms such as (a)~the DataSynthesizer~\cite{ping2017datasynthesizer}, (b)~the Synthetic Data Vault~\cite{patki2016synthetic},
and (c) the Howso engine~\cite{hazard2019natively}.

To structure this inquiry, we ask these questions: \\
\begin{itemize}
    \item RQ1: When considering the privacy, which synthetic data generation algorithm can generate the synthetic data with the highest privacy preservation score?
    \item RQ2: Which synthetic data generation algorithm can generate data that has higher similarity to the original data?
    \item RQ3: When the machine learning model is trained on the synthetic data, can the model achieve compatible performance with those trained on the original data?
    \item RQ4: Which algorithm has the best scalability?
    \item RQ5: What suggestions can we provide from analyzing the conclusions in RQ1 to RQ4?
\end{itemize}

The contributions of this paper are
\begin{itemize}
    \item We proposed a random projection based synthetic data generation framework.
    \item We made an empirical experiment to compare our proposed method, Howso engine, and two state-of-the-art synthetic data generation algorithms in different aspects such as (a) privacy preservation, (b) statistical measurements, (c) marginal probability, and (d) performance on training the machine learning models.
    \item We found random projection based framework and Howso engine outperform state-of-the-art methods in some of the aspects.
    \item The random projection based framework, in terms of scalability, can run significantly faster than Howso engine, and have similar runtime comparing to state-of-the-art methods, while outperform on more metrics than state-of-the-art methods.
\end{itemize}
The rest of this paper is constructed as follow: Section~\ref{background} illustrates the background of synthetic data generation, privacy \& accuracy, and literature review on the synthetic data generation. Section~\ref{howso} presents the Howso engine. Section~\ref{randomprojection} illustrates our proposed random projection based synthetic data generation framework. Section~\ref{methods} shows two state-of-the-art synthetic data generation algorithms DataSynthesizer and Synthetic Data Vault. Section~\ref{setup} presents the summary of benchmarks, evaluation metrics, and statistical analysis used in our experiment. Section~\ref{results} shows our experimental results and our analysis to the results. We also discuss the threat to validity of our experiment in Section~\ref{threat}, and make a conclusion to this study in Section~\ref{conclusion}.

\begin{table*}[]
    \centering
    \begin{tabular}{m{4.5cm}|m{0.6cm}|m{3cm}|m{1.6cm}|m{4cm}|m{1cm}}
        \textbf{Paper} & \textbf{Year} & \textbf{Algorithm(s)} & \textbf{Technique} & \textbf{Metric(s)} & \textbf{Region} \\
        \hline
        \hline
        \multicolumn{6}{c}{Synthetic Data Generation} \\
        \hline
        \hline
        \rowcolor{gray!20}
        The Synthetic Data Vault~\cite{patki2016synthetic} & 2016 & SDV & Gaussian Copula & Accuracy \& Qualititive findings & N/A \\
        \rowcolor{white}
        synthpop: Bespoke creation of synthetic data in R~\cite{nowok2016synthpop} & 2016 & SynthPop & CART tree based method & statistical metrics & N/A \\
        \rowcolor{gray!20}
        DataSynthesizer: Privacy-Preserving Synthetic Datasets~\cite{ping2017datasynthesizer} & 2017 & DataSynthesizer & Greedy Bayes & Similarity measure (e.g feature distribution) & Urban science \\
        \rowcolor{white}
        RPA and L-System Based Synthetic Data Generator for Cost-efficient Deep Learning Model Training~\cite{fiestas2021rpa} & 2021 & Lindenmayer Systems & RPA \& L-system & Machine learning metrics & Image \\
        \rowcolor{gray!20}
        Synthetic data generation using building information models~\cite{hong2021synthetic} & 2021 & CycleGAN & GAN & Average precision & Image \\
        \rowcolor{white}
        Fedsyn: Synthetic data generation using federated learning~\cite{behera2022fedsyn} & 2022 & FedSyn & Federated learning & Accuracy \& subjective analysis & Image \\
        \rowcolor{gray!20}
        Generation of synthetic tympanic membrane images...~\cite{suresh2023generation} & 2023 & GAN & GAN & Human review & Medical \\
        \hline
        \hline
        \multicolumn{5}{c}{Utility Validation} \\
        \hline
        \hline
        \rowcolor{white}
        Utility of synthetic microdata generated using tree-based methods~\cite{nowok2015utility} & 2015 & SynthPop & N/A & Propensity score, Coefficient estimates, Mean overlap in the 95\% confidence intervals & N/A \\
        \rowcolor{gray!20}
        The validity of synthetic clinical data: a validation study of...~\cite{chen2019validity} & 2019 & Synthea & N/A & Medical quality measurement & Medical \\
        \rowcolor{white}
        On the Utility of Synthetic Data: An Empirical Evaluation on Machine Learning Tasks~\cite{hittmeir2019utility} & 2019 & DS, SDV & N/A & Distribution, Correlation coefficient, Distance of nearest neighbors, Accuracy & N/A \\
        \rowcolor{gray!20}
        Empirical evaluation on synthetic data generation with generative adversarial network~\cite{lu2019empirical} & 2019 & GAN & N/A & Correlation metrix, Accuracy, Privacy metrics & N/A \\
        \rowcolor{white}
        Generation and evaluation of synthetic patient data~\cite{goncalves2020generation} & 2020 & Sampling from marginal, Bayansian network, GAN, Gaussian process & N/A & LK divergence, pairwise correlation difference, log-cluster, cross-classifcation & Medical \\
        \rowcolor{gray!20}
        Can synthetic data be a proxy for real clinical trial data? A validation study~\cite{azizi2021can} & 2021 & Sequential decision tree & N/A & Bivariate analysis, multivariate analysis & Medical \\
        \rowcolor{white}
        Fake It Till You Make It: Guidelines for Effective Synthetic Data Generation~\cite{dankar2021fake} & 2021 & DS, SDV, Synthpop & N/A & Propensity score pMSE, Accuracy & N/A \\
        \rowcolor{gray!20}
        Generating and evaluating cross‐sectional synthetic electronic healthcare data...~\cite{wang2021generating} & 2021 & N/A & N/A & Univariate/Multivariate distance, Privacy preservation & Medical \\
        \rowcolor{white}
        Synthetic data use: exploring use cases to optimise data utility~\cite{james2021synthetic} & 2021 & N/A & N/A & Distribution comparison, Hellinger distance, Accuracy, Bivariate correlation, Area under the receiver operating characteristic & N/A \\
        \rowcolor{gray!20}
        Utility Metrics for Evaluating Synthetic Health Data Generation Methods: Validation Study~\cite{el2022utility} & 2022 & N/A & N/A & Maximum mean discrepancy, Hellinger distance, Wasserstein distance & Medical \\
        \rowcolor{white}
        A multi-dimensional evaluation of synthetic data generators~\cite{dankar2022multi} & 2022 & N/A & N/A & Attribute fidelity, Bivariate fidelity, Population difelity, Application fidelity & N/A \\
        \rowcolor{gray!20}
        How Faithful is your Synthetic Data? Sample-level Metrics for Evaluating and Auditing Generative Models~\cite{alaa2022faithful} & 2022 & GAN, VAE, WGAN-GP, ADS-GAN & N/A & Alpha-precision, Beta-recall, Authenticity & N/A \\
    \end{tabular}
    \caption{Literature review on recent synthetic data studies. The literature can be described in two categories: (a) The design of new synthetic data generation algorithms (Synthetic Data Generation), and (b) The implementation of synthetic data generation algorithms and the evaluation on different utility metrics (Utility Validation). Some of long titles are not fully shown in the first column. For the entire title please refer to the specific reference.}
    \label{tab:literature}
\end{table*}

\section{Background}\label{background}
\subsection{Synthetic Data Generation}
Synthetic data is used to substitute the real data which cannot be shared to public due to privacy concerns. It is usually generated by the generative model which learns the patterns of data from the original dataset. More specifically, the data synthesize process usually can be described as follow: \\
\textbf{Given a dataset $\mathcal{D}$ which some of the features $\{ f_i, f_j, \cdots, f_k \}$ have privacy concerns, the generative model $\mathcal{M}$ learns the statistical properties $p$ for each feature and the correlation $c$ between different features from the original dataset $\mathcal{D}$. Then the generative model $\mathcal{M}(p, c)$ will generate the synthetic dataset $\mathcal{D}'$ that similar to $\mathcal{D}$, but no original identity can be detected by the synthetic value in the features $\{ f_i, f_j, \cdots, f_k \}$.}

Researchers mainly focus on the (a) privacy preservation~\cite{liu2022privacy, cunningham2021privacy, mayer2020privacy} and (b) model accuracy~\cite{jeske2005generation, boutros2022sface} to judge if the synthesized data is both informative and safe to share. As we will talk later in this paper, we evaluate the synthetic data in our study through both privacy preservation score and model performance score, as well as the statistical distribution score.

\subsection{Related Work}
Synthetic data generation and evaluation has been widely studied in the past few years. The literature mainly split in two directions. One direction is the development of  new synthetic data generation algorithms such as~\cite{ping2017datasynthesizer, patki2016synthetic}. Another direction is applying well-established generation methods to different datasets in various domains, and evaluating the results through different metrics~\cite{el2022utility}.

We search the literature in Google Scholar for what has  been published in  top venues\footnote{As defined by Google Scholar metrics.} then summarize the related work in Table~\ref{tab:literature}. As our best knowledge, recent literature on synthetic data can be mainly split into two regions. The studies in the first category develops the new generation algorithm, and the studies in the second category mainly evaluates the utility of synthetic data generated by different models through different metrics. In our study, we introduce Howso engine   and our proposed random projection based synthetic data generation framework, so we compare these two methods to the algorithms proposed in the literature in the first category. We choose our comparison objects with following rules:
\begin{itemize}
    \item First, the study goal should focus on the tabular data since all benchmarks in our study are tabular based.
    \item Second, the implementation should be based on Python to compatible with our inputs.
\end{itemize}
This meant we focused on  (a)~the DataSynthesizer~\cite{ping2017datasynthesizer} and (b)~ the Synthetic Data Vault~\cite{patki2016synthetic}. Hence, in our study, we compare Howso engine and random projection based framework to DataSynthesizer and Synthetic Data Vault.


\section{Howso Engine}\label{howso}
Howso Engine is developed by Howso\footnote{https://www.howso.com/}. It is an AI engine which supports the synthetic data generation. Specifically, Howso Engine utilizes the k-nearest neighbors to synthesize data with both global and local distributions~\cite{banerjee2023surprisal}. The algorithm contains three parameters to control the search, and the best combination of the parameters is found by the grid search. Since there is only two parameters that need to be explored, the grid search is very fast.
\begin{itemize}
    \item The Minkowski coefficient $p \in [ 0.1, 2.0 ]$ which controls the calculation of distance.
        \begin{equation}
            d(x, y, \Delta) = (\sum_{i=1}^{n} \Delta(x_i, y_i)^p)^{1/p}
        \end{equation}
    \item Iteration parameter $l = 6$ which is used to find the parameter for distance calculation.
    \item The number of neighbors $k \in [5, 22]$ during the execution of $k$-nearest neighbors algorithm.
\end{itemize}

In the distance calculation, $\Delta$ is the specific function to measure the distance between two variables. Howso Engine implements the Lukaszyk-Karmowski metric (LK metric) for the Laplace distribution~\cite{hazard2019natively}. Specifically,
\begin{equation}
    \Delta(x_i,y_i) = |x_j - x_i| + \frac{1}{2} e^{-\frac{|x_j - x_i|}{b}} \left( 3b+|x_j - x_i| \right)
\end{equation}
The Laplace distribution is preferred here since it makes entropy-minimizing assumptions about the underlying data, and more performant than Gaussian distribution. In the above calculation, $b$ is the surprisal that needs to be found through multiple iterations for each dataset. Specifically, the initial $b$ is set to $1/k$ where $k$ is the number of neighbors used in $k$-nearest neighbor algorithm. The first iteration finds the local neighbors by traditional Minkowski distance since no information in the initial round. In the end of the initial round, we can update the surprisal $b$:
\begin{itemize}
    \item For the numerical feature $k$, Howso Engine uses mean absolute deviation (MAE):
    \begin{equation}
        b = \frac{1}{n} \sum_{i=1}^{n} |x_i - \mu_k|
    \end{equation}
    \item For the symbolic feature, Howso Engine uses mode and accuracy instead of mean and MAE.
\end{itemize}

Once the value of $b$ is stabilized (which we found $b$ tends to stable in 6 iterations), Howso Engine iteratively picks one random feature to synthesize. Specifically,
\begin{itemize}
    \item The values in the first picked feature will be synthesized on the basis of the global histogram (for nominal features) or the global Laplace distribution (for continuous features).
    \item For all subsequent features, values are generated based on the distribution in the $k$ nearest neighbors to the partially synthesized cases.
\end{itemize}
The above process is done $m$ times per feature where $m$ is the number of synthetic instances that request to be generated.

\section{Recursive Random Projection}\label{randomprojection}
Recursive random projection, as its name shows, projects the high-dimensional data into different low-dimensional clusters by using the random pivot selection procedure recursively~\cite{vempala2005random} . In the synthetic data generation, the relationships between different features are hard to capture. Previous literature uses causal graph~\cite{ping2017datasynthesizer} or covariance metrics~\cite{patki2016synthetic} to explore the connections between different features. In this study, we adopt the random projection to split the data into different clusters, which each cluster capture the data points that have similar feature patterns. In this section, we will introduce our design of random projection framework from the following two aspect:
\begin{itemize}
    \item First, the \textbf{cluster} algorithm which split the data into different clusters.
    \item Second, the \textbf{mutation \& crossover} operators which mutate the data points in the same cluster to generate synthetic data points that have strong connection to the original data points.
\end{itemize}

The \textbf{cluster} algorithm aims to split the data points into different clusters, which each cluster will contain data points with similar patterns. To achieve that, we utilize the FASTMAP random projection algorithm~\cite{faloutsos1995fastmap}. With a set of data points, FASTMAP uses the cosine rule to project the data points into the hyperplane formed by two farest points. More specifically, with two farest points $a$ and $b$, any third point $c$ from the set of data points can be mapped into the line connecting $a$ and $b$ by
\begin{equation}
    x = (a^2 + c^2 - b^2) / (2c)
\end{equation}
Algorithm~\ref{alg:cluster} shows the recursive clustering procedure, which in each function call, the algorithm first check if the number of current candidates in $\mathcal{C}$ less than the threshold $t$ (line 1). If the number does not reach the threshold, then stop the recursive. Otherwise, the \textbf{split} function will bi-cluster the current candidates, and perform the \textbf{cluster} algorithm again to the two sub-clusters (line 2-5). Algorithm~\ref{alg:split} illustrates in detail how the \textbf{split} function works. More specifically, it firstly picks a random pivot, and finds two farest points based on the random pivot (line 1-4). After that, as we stated above, it uses the cosine rule to map all other data points to the line formed by two farest point (line 5-10). Finally, it returns two subsets based on the distance calculated by cosine rule (line 11-13).

\begin{algorithm}[t]
    \caption{\textbf{cluster}: The overall recursive clustering structure inside the random projection. It outputs a tree structure $\mathcal{T}$ that the nodes in the $i^{th}$ depth are bi-clustered into the $(i+1)^{th}$ depth.}
    \label{alg:cluster}
    \small
    \begin{algorithmic}[1]
        \Require $\mathcal{C}$, $t$, $\mathcal{T}$, $\mathcal{N}$
        \If{$|\mathcal{C}| > t$}
        \State $\mathcal{C}_{east}$, $\mathcal{C}_{west}$ = \textbf{split}($\mathcal{C}$) 
        \Comment{bi-cluster}
        \State $\mathcal{N}$.left.\textbf{value}, $\mathcal{N}$.right.\textbf{value} = $\mathcal{C}_{east}$, $\mathcal{C}_{west}$
        \State \textbf{cluster}($\mathcal{C}_{east}$, $t$, $\mathcal{T}$, $\mathcal{N}$.left.\textbf{value})
        \Comment{left node recursive}
        \State \textbf{cluster}($\mathcal{C}_{west}$, $t$, $\mathcal{T}$, $\mathcal{N}$.right.\textbf{value})
        \Comment{right node recursive}
        \EndIf
    \end{algorithmic}
\end{algorithm}
\begin{algorithm}[t]
    \caption{\textbf{split}: Split a set of candidates $\mathcal{C}$ into two subsets by using the FASTMAP technology~\cite{faloutsos1995fastmap, platt2005fastmap}.}
    \label{alg:split}
    \begin{algorithmic}[1]
        \Require $\mathcal{D}$
        \State $rand$ = \textbf{random}$(1, |\mathcal{D}|)$
        \State $pivot = \mathcal{D}(rand)$ 
        \Comment{pick a random point as pivot} 
        \State $p_E$ = \textbf{mostDistance}($\mathcal{D}$, $pivot$) 
        \Comment{farthest point to pivot}
        \State $p_W$ = \textbf{mostDistance}($\mathcal{D}$, $p_E$) 
        \Comment{farthest point to east}
        \State $c$ = \textbf{distance}($p_E$, $p_W$)
        \Comment{Similarity measure as distance}
        \For{$idx = 1:|\mathcal{D}|$}
            \State $a$ = \textbf{distance}($\mathcal{D}$($idx$), $p_E$)
            \State $b$ = \textbf{distance}($\mathcal{D}$($idx$), $p_W$)
            \State $\mathcal{D}(idx)$.d = ($a^2$ + $c^2$ - $b^2$) / (2$c$)
            \Comment{cosine rule}
        \EndFor
        \State sorted = \textbf{sort}($\mathcal{D}$.d) 
        \Comment{Sort all points via distance}
        \State $\mathcal{D}_E$ = $\mathcal{D}$[:0.5*\textbf{size}(sorted)]
        \State $\mathcal{D}_W$ = $\mathcal{D}$[0.5*\textbf{size}(sorted):]
    \end{algorithmic}
\end{algorithm}

Traditional clustering algorithms require $O(N^2)$ calculations to fully split the data. However, random projection can achieve that in $O(2N)$, which is much faster than traditional clustering algorithms. This is the intuition we use random projection to reduce the scalability on generating synthetic data.

After the random projection returns the clusters, we then utilize the \textbf{mutation and crossover} operators used in the differential evolution algorithm to generate the synthetic data~\cite{storn1997differential}. In the differential evolution algorithm, a better solution can be found in the current region of data points by mutating the data points as follow
\begin{equation}
    y_{\text{new}_i} = 
    \begin{cases}
        x_{1_i} + F * (x_{2_i} - x_{3_i}), & \text{if } r_i < CR \text{ or } k = R \\
        x_{\text{old}_i}, & \text{O.W.}
    \end{cases}
\end{equation}
where $x_{\text{old}_i}$ is the point from the original set of candidates, and $x_1$, $x_2$, and $x_3$ are three other random points from the set of candidates. $F$ is the difference scaling factor from 0 to 1, and $CR$ is the crossover probability that also from 0 to 1. Large $F$ indicates more scaling on the difference of two data points during the mutation, and large $CR$ indicates more probability the new candidate has a new value in each index. $R$ is a random index such that the value in that index must be mutated. This can prevent the duplicated new candidate when $CR$ is very small. We hypothesis that the synthetic data generated by this mutation and crossover operator can capture the feature information from the original data in each cluster since each cluster includes the data points that are close to each other.

\section{Experimental Methods}\label{methods}
In this section, we will briefly explain two state-of-the-art synthetic data generation algorithms we compared to.

\subsection{DataSynthesizer}
DataSynthesizer is proposed by Ping et al.~\cite{ping2017datasynthesizer}. Their framework can handle two different attribute mode. One is \textit{independent attribute mode}, which each feature is treated individually. Another one is \textit{correlated attribute mode}, which the causal graph is used to describe the relationships between each feature. First of all, DataSynthesizer implements a module called DataDescirber to capture the data type for each feature, as well as its distribution and correlation. Also, it will add the noise to the data distribution to preserve the privacy. Secondly, DataGenerator module will generate the synthetic data based on the attribute mode.

More specifically, for the \textit{independent attribute mode}, DataDescriber performs the frequency-based estimation of the unconditioned probability distributions of each attribute~\cite{ping2017datasynthesizer}. To preserve the privacy, the noise $Lap(\frac{1}{n \epsilon})$ will be added to the distribution where $n$ is the size of the inputs and $\epsilon$ is set to 0.1 by default. DataGenerator will then uses the distribution to generate the synthetic data for each feature.

On the other hand, the \textit{correlated attribute mode} is quite different to the \textit{independent attribute mode}. The GreedyBayes algorithm is utilized to construct the causal graph for all the features. GreedyBayes is a kind of greedy selection algorithm that select the highest correlated feature which maximize the mutual information to the subsets of features that have been visited. The noise $Lap(\frac{4(|\mathcal{A}|-k)}{n \cdot \epsilon})$ will also be added to preserve the privacy. After the causal graph is generated, the algorithm will then use the knowledge from causal graph and the distribution to generate the synthetic data.

\subsection{Synthetic Data Vault}
Synthetic Data Vault (SDV) is proposed by Patki et al.~\cite{patki2016synthetic}. The basic intuition behind SDV is to use the Gaussian Copula as the generative model to synthesize the data based on the distribution and the covariance of the features. More specifically, with a dataset that has columns $\{ c_1, c_2, \cdots, c_n \}$, we use $\{ f_1, f_2, \cdots, f_n \}$ to express the cumulative distribution function for those columns. The Gaussian Copula will calculate the inverse cumulative distribution functions $\phi$ of the Gaussian distribution applied to the original cumulative distribution function $\{ f_1, f_2, \cdots, f_n \}$. Algorithm~\ref{alg:GaussianCopula} shows how to use the Gaussian Copula to calculate the covariance matrix when given the dataset.

\begin{algorithm}[]
    \caption{\textbf{Gaussian Copula}: The Gaussian Copula algorithm for analyzing the distribution and covariance of the dataset~\cite{patki2016synthetic}. Return the covariance matrix $\Sigma$.}
    \label{alg:GaussianCopula}
    \small
    \begin{algorithmic}[1]
        \Require $\mathcal{D} = \{d_1, d_2, \cdots, d_p \}$, $\mathcal{F} = \{f_1, f_2, \cdots ,f_n\}$
        \For{$d_i$ in $\mathcal{D}$}
            \State $Y_i = [\Phi^{-1}(f_0(d_{i0})), \Phi^{-1}(f_1(d_{i1})), \cdots, \Phi^{-1}(f_n(d_{in}))]$
        \EndFor
        \State $\Sigma = \textbf{computeCovariance}(\{Y_1, Y_2, \cdots, Y_n\})$
    \end{algorithmic}
\end{algorithm}

To generate synthetic data, for a given row, SDV will generate the synthetic value based on the feature distributions. Moreover, if there is information on the other features, then the covariance information calculated from Gaussian Copula will also be used along with the feature distribution to generate the synthetic data.

We use the SDV public API\footnote{https://github.com/sdv-dev/SDV/tree/main} to implement the SDV in our study. Note that their online API also includes the Machine Learning based generative model and the GAN based generative model. Hence, in our study, we compare our methods to both three SDV generative models.

\begin{table}[!h]
    \centering
    \begin{tabular}{c|c|c|c}
        \textbf{Benchmark} & \textbf{\# Rows} & \textbf{\# Cols} & \textbf{Task} \\
        \hline
        glass & 203 & 10 & multi-classification \\
        596\_fri\_c2\_250 & 250 & 6 & regression \\
        breast\_cancer & 286 & 10 & classification \\
        cars & 392 & 9 & multi-classification \\
        irish & 500 & 6 & classification \\
        522\_pm10 & 500 & 8 & regression \\
        profb & 673 & 10 & classification \\
        tic\_tac\_toe & 958 & 10 & classification \\
        churn & 5000 & 21 & classification \\
        adult & 48842 & 15 & classification
    \end{tabular}
    \caption{Summary of benchmarks used in our experiment.}
    \label{tab:benchmark}
\end{table}

\section{Experimental Setup}\label{setup}
In this section we will illustrate following things: (a) the benchmarks, (b) the evaluation metrics, and (c) the statistical analysis procedure.

\subsection{Benchmark}
Table~\ref{tab:benchmark} shows   10 machine learning datasets from the Penn Machine Learning Benchmark (PMLB) datasets\footnote{https://github.com/EpistasisLab/pmlb}~\cite{olson2017pmlb}. To select those ten datasets, we firstly counted the number of instances in all datasets in PMLB. After that, we group those datasets in the 10 clusters based on the number of instances, and randomly pick one dataset from each cluster. This step can ensure that the benchmarks used in our experiment have different size, and thus we can analyze the runtime of each synthetic data generator more empirically. Moreover, we manually inspect the selected datasets, and replace some of them to the one with more privacy concerns. Please note that the ``privacy concerns'' means some features are very informative to identify the individuals. In this case, the datasets with those features will not be able to share in the real world. As we stated in the Introduction section, the synthetic data is designed to replace those datasets with sensitive information, and that is the reason we will make this replacement to our selected benchmarks. However, some datasets we used may not have sensitive features, and there is no equivalent one to replace with, then we will keep using these datasets and assume they have the privacy concerns as other datasets have. The last point is that we also keep some datasets which the task is either multi-class classification or regression, to compare the synthetic data generation performance in both regression and classification task. The summary of our selected benchmarks is shown in Table~\ref{tab:benchmark}.

\subsection{Evaluation Metrics}\label{metric}
The \textbf{Metric(s)} column in Table~\ref{tab:literature} shows different metrics have been used in the past literature. In the general application of synthetic data, four types of metrics are highly used.
\begin{itemize}
    \item \textbf{Privacy metric} which evaluates the privacy preservation of the synthetic data.
    \item \textbf{Informational coefficients metric} which evaluates the statistical information of the synthetic data.
    \item \textbf{Distribution metric} which evaluates the synthetic data through joint distribution.
    \item \textbf{Model performance metric} which build the machine learning models on the original data and the synthetic data and test their performance on the original test data.
\end{itemize}
In our study, we evaluate each algorithm through these four metrics. The details of each metric is explained in the following subsections. All these metrics are used  to validate the synthetic data.

\subsubsection{Privacy Preservation}\label{privacypreservation}
Privacy preservation evaluates the privacy of synthetic data. It checks if the distance from a synthetic data point to the density of the region of its nearest original neighbor is small or not. More specifically, to evaluate the privacy preservation score of a synthetic data point $x_{syn}$, we first find its nearest original neighbor $x_{ori}$, and calculate the distance from $x_{syn}$ to $x_{ori}$ as $d$. After that, we find $k$-nearest original neighbors of $x_{ori}$, and calculate the minimum distance $d_{min}$ between any two of the points in the group of that $k$-nearest neighbors. The final score of privacy preservation is the minimum distance ratio of distance $d$ and the distance $d_{min}$. Though larger score indicates better privacy preservation, the minimum distance ratio of 1 already indicates that the synthetic point is located outside of the density region of its nearest original neighbor. Hence any value greater or equal than 1 can indicate a good privacy preservation.

\subsubsection{Statistical Similarity}\label{statisticalsimilarity}
Statistical similarity compares the synthetic data to the original data through statistical measurements. The statistical measurements are split into three parts.
\begin{itemize}
    \item \textbf{Central tendency} which includes \textit{mean, mode, median, 25th percentile, 75th percentile, minimum, and maximum}.
    \item \textbf{Variability of dispersion} which includes \textit{entropy, Kurtosis, mean absolution deviation, standard deviation, skew, and variance}.
    \item \textbf{Frequency distribution} which describes the \textit{uniqueness} of the data.
\end{itemize}

All those metrics are calculated in both the synthetic dataset and the original dataset. For all pairs of score in $\{S^{ori}, S^{syn}\}$ for a certain feature, we calculate the \textbf{SMAPE}. Specifically, SMAPE is the symmetric mean absolute percentage error which is an accuracy measure based on relative error~\cite{shcherbakov2013survey}, and is calculated as follow
\begin{equation}\label{SMAPE}
    \text{SMAPE} = \frac{1}{n}\sum_{t=1}^{n}\frac{|s_t^{syn} - s_t^{ori}|}{(|s_t^{ori}| + |s_t^{syn}|) / 2}
\end{equation}
where $s_t^{syn}$ and $s_t^{ori}$ are the scores of statistical measurement $t$ in synthetic data $S^{syn}$ and original data $S^{ori}$. The final statistical similarity score is the average SMAPE of all features. The smaller score in this metric means less difference between synthetic data and original data on the statistical measurements.

\subsubsection{Marginal Distribution Similarity}
Marginal distribution similarity evaluates the synthetic data through the marginal distribution. The marginal distribution of the numeric feature is estimated through the KNN density estimation and the marginal distribution of the nominal feature is estimated using normalized value counts. The score of marginal distribution similarity is then calculated by comparing the estimated distributions using \textit{Jensen-Shannon divergence}. Specifically, Jensen-Shannon divergence measures the similarity between two probability distribution~\cite{fuglede2004jensen}, and is calculated as follow
\begin{equation}\label{JSdivergence}
    \delta_{JS}(\mathcal{P}, \mathcal{Q}) = \delta_{KL}(\mathcal{P} || \mathcal{M}) + \delta_{KL}(\mathcal{Q} || \mathcal{M})
\end{equation}
where $\mathcal{M} = (\mathcal{P} + \mathcal{Q}) / 2$. In the above formula, $\delta_{KL}$ is the Kullback-Leibler divergence~\cite{csiszar1975divergence} which measures the distance of probability distribution $P$ to the reference probability distribution $Q$ by
\begin{equation}
    \delta_{KL}(\mathcal{P} || \mathcal{Q}) = \int_{-\infty}^{\infty} p(x) \log{(\frac{p(x)}{q(x)})} dx
\end{equation}
The smaller Jensen-Shannon divergence value indicates that two probability distributions are similar.

\subsubsection{Model Comparison}
Model comparison evaluates the quality of synthetic data on training the machine learning models. More specifically, both the original data and synthetic data is split into 80\% of training set and 20\% of test set. After that, the LightGBM classification model or regression model is trained on the \textbf{synthetic training set}, and test on the \textbf{original test set}. For the regression task, we report \textit{RMSE}, \textit{R$^2$}, and \textit{Spearman correlation coefficient}. Specifically
\begin{itemize}
    \item \textbf{RMSE} is the root mean square error which measures the average difference between the predicted values from a regression model and the actual values~\cite{chai2014root}. It is calculated as follow
        \begin{equation}\label{RMSE}
            \text{RMSE} = \sqrt{\frac{\sum_{i=1}^{N}(\hat{y}_i - y_i)}{N}}
        \end{equation}
    \item \textbf{R$^2$} (or say coefficient of determination) represents how well the data fit the regression model~\cite{gelman2019r}. Specifically, let $\overline{y}$ be the mean of all observation $y$, and $\hat{y}$ be the predicted value from the model, R$^2$ is calculated as follow
        \begin{equation}\label{R2}
            \text{R}^2 = 1 - \frac{\sum_{i=1}^N (y_i - \hat{y}_i)^2}{\sum_{i=1}^N (y_i - \overline{y})^2}
        \end{equation}
    \item \textbf{Spearman correlation coefficient} is the statistical measure which check the linear correlation between two populations~\cite{hauke2011comparison}. Given a pair of same feature from original dataset and synthetic dataset $X_{ori}$ and $X_{syn}$, the score of Spearman is calculated by
        \begin{equation}\label{Spearman}
            \rho_{X_{ori}, X_{syn}} = \frac{\mathbb{E}((X_{ori} - \mu_{X_{ori}})(X_{syn} - \mu_{X_{syn}}))}{\sigma(X_{ori})\sigma(X_{syn})}
        \end{equation}
    where $\mu$ is the mean and $\sigma$ is the standard deviation of the population. The overall score of Spearman is the geometric mean of all features. We expect higher Spearman value since score 1 means two populations perfectly fit the linear correlation.
\end{itemize}
And for the classification task, we report \textit{accuracy}, \textit{precision}, \textit{recall}, and \textit{Matthews correlation coefficient}. Specifically, in the classification task, we notate $TP$, $TN$, $FP$, $FN$ as the value of \textit{true positive, true negative, false positive,} and \textit{false negative} returned from the confusion matrix,
\begin{itemize}
    \item \textbf{Accuracy} evaluates the ratio of number of \textit{correct predictions} over the \textit{total number of predictions}.
        \begin{equation}
            \text{Accuracy} = \frac{TP + TN}{TP + FP + FN + TN}
        \end{equation}
    \item \textbf{Precision} evaluates the ratio of number of \textit{correct positive predictions} over the total number of \textit{predicted positive cases}.
        \begin{equation}
            \text{Precision} = \frac{TP}{TP + FP}
        \end{equation}
    \item \textbf{Recall} evaluates the ratio of number of \textit{correct positive predictions} over the total number of \textit{actual positive cases}.
        \begin{equation}
            \text{Recall} = \frac{TP}{TP + FN}
        \end{equation}
    \item \textbf{Matthews correlation coefficient} evaluates the prediction performance by summarizing the entire confusion matrix.
        \begin{equation}
            \text{MCC} = \frac{TN * TP - FN * FP}{\sqrt{(TP + FP)(TP + FN)(TN + FP)(TN + FN)}}
        \end{equation}
\end{itemize}
For all those classification metrics, the value close to 1 will indicate the better performance.

\begin{table*}[t]
    \centering
    \begin{tabular}{c||c|c|c|c|c|c|c|c|c|c||c}
        \multirow{3}{*}{\textbf{Algorithm}} & \multicolumn{10}{c||}{\textbf{Benchmarks}} & \multirow{3}{*}{\textbf{Wins}} \\
        \cline{2-11}
         & \multirow{2}{*}{glass} & 596\_fri\_ & breast\_ & \multirow{2}{*}{cars} & \multirow{2}{*}{irish} & 522\_ & \multirow{2}{*}{profb} & tic\_tac\_ & \multirow{2}{*}{churn} & \multirow{2}{*}{adult} & \\
         & & c2\_250 & cancer & & & pm10 & & toe & & & \\
        \hline
        DataSynthesizer-Ind & \cellcolor{gray!30}1.48 & 0.53 & \cellcolor{gray!30}1.05 & \cellcolor{gray!30}3.95 & \cellcolor{gray!30}6.00 & \cellcolor{gray!30}1.34 & \cellcolor{gray!30}1.10 & \cellcolor{gray!30}0.00 & \cellcolor{gray!30}1.89 & \cellcolor{gray!30}17.31 & 9 \\
        DataSynthesizer-Cor & \cellcolor{gray!30}2.24 & \cellcolor{gray!30}1.22 & \cellcolor{gray!30}2.00 & \cellcolor{gray!30}3.78 & \cellcolor{gray!30}1.00 & \cellcolor{gray!30}1.48 & \cellcolor{gray!30}1.07 & \cellcolor{gray!30}0.00 & \cellcolor{gray!30}1.91 & \cellcolor{gray!30}1.68 & 10 \\
        Synthetic Data Vault-ML & 0.88 & 0.60 & 0.00 & \cellcolor{gray!30}1.17 & 0.15 & 0.80 & 0.50 & \cellcolor{gray!30}0.00 & 0.57 & 0.44 & 2 \\
        Synthetic Data Vault-GC & \cellcolor{gray!30}1.44 & 0.64 & 0.00 & 0.89 & 0.00 & 0.53 & 0.33 & \cellcolor{gray!30}0.00 & \cellcolor{gray!30}1.71 & 0.38 & 3 \\
        Synthetic Data Vault-GAN & \cellcolor{gray!30}1.44 & 0.74 & 0.00 & \cellcolor{gray!30}1.68 & 0.00 & 0.80 & 0.54 & \cellcolor{gray!30}0.00 & 0.76 & 0.00 & 3 \\ 
        Recursive Random Projection & 0.45 & 0.46 & 0.00 & 0.29 & 0.00 & 0.56 & 0.55 & \cellcolor{gray!30}0.00 & 0.70 & 0.00 & 1 \\
        Howso Engine & \cellcolor{gray!30}1.00 & \cellcolor{gray!30}1.00 & 0.00 & \cellcolor{gray!30}1.00 & \cellcolor{gray!30}1.25 & \cellcolor{gray!30}1.00 & 0.45 & \cellcolor{gray!30}0.00 & 0.82 & 0.01 & 6 \\
    \end{tabular}
    \caption{Privacy preservation score. Higher values are better. The dark gray cell marks the algorithm in the highest rank by Scott-Knott analysis, and the light gray cell marks the algorithm in the second highest rank.}
    \label{tab:anonymityPreservation}
\end{table*}

\begin{table*}[t]
    \centering
    \begin{tabular}{c||c|c|c|c|c|c|c|c|c|c||c}
        \multirow{3}{*}{\textbf{Algorithm}} & \multicolumn{10}{c||}{\textbf{Benchmarks}} & \multirow{3}{*}{\textbf{Wins}} \\
        \cline{2-11}
         & \multirow{2}{*}{glass} & 596\_fri\_ & breast\_ & \multirow{2}{*}{cars} & \multirow{2}{*}{irish} & 522\_ & \multirow{2}{*}{profb} & tic\_tac\_ & \multirow{2}{*}{churn} & \multirow{2}{*}{adult} & \\
         & & c2\_250 & cancer & & & pm10 & & toe & & & \\
        \hline
        DataSynthesizer-Ind & 0.77 & 0.84 & 0.50 & 0.35 & 0.55 & 0.50 & 0.56 & 0.40 & 0.60 & 0.43 & 0 \\
        DataSynthesizer-Cor & 0.74 & 0.75 & 0.14 & 0.35 & 0.44 & 0.55 & 0.45 & 0.17 & 0.66 & 0.44 & 0 \\
        Synthetic Data Vault-ML & 0.40 & 0.51 & \cellcolor{gray!30}0.10 & 0.22 & \cellcolor{gray!30}0.07 & 0.39 & 0.32 & 0.17 & 0.32 & 0.33 & 2 \\
        Synthetic Data Vault-GC & 0.45 & \cellcolor{gray!30}0.40 & \cellcolor{gray!30}0.10 & 0.13 & 0.15 & 0.47 & \cellcolor{gray!30}0.13 & \cellcolor{gray!30}0.02 & 0.32 & 0.32 & 4 \\
        Synthetic Data Vault-GAN & 0.32 & 0.66 & 0.12 & 0.22 & 0.09 & 0.45 & 0.25 & 0.05 & 0.20 & 0.16 & 0 \\ 
        Recursive Random Projection & \cellcolor{gray!30}0.13 & \cellcolor{gray!30}0.38 & 0.12 & 0.12 & 0.17 & 0.29 & 0.26 & \cellcolor{gray!30}0.00 & 0.32 & \cellcolor{gray!30}0.11 & 4 \\
        Howso Engine & 0.16 & 0.45 & \cellcolor{gray!30}0.06 & \cellcolor{gray!30}0.08 & 0.16 & \cellcolor{gray!30}0.26 & 0.15 & \cellcolor{gray!30}0.00 & \cellcolor{gray!30}0.19 & 0.13 & 5 \\
    \end{tabular}
    \caption{Statistical similarity score. The overall score is evaluated through SMAPE in equation~\ref{SMAPE}. Smaller values are better. The dark gray cell marks the algorithm in the highest rank by Scott-Knott analysis, and the light gray cell marks the algorithm in the second highest rank.}
    \label{tab:descriptiveStatistics}
\end{table*}

\begin{table*}[t]
    \centering
    \begin{tabular}{c||c|c|c|c|c|c|c|c|c|c||c}
        \multirow{3}{*}{\textbf{Algorithm}} & \multicolumn{10}{c||}{\textbf{Benchmarks}} & \multirow{3}{*}{\textbf{Wins}} \\
        \cline{2-11}
         & \multirow{2}{*}{glass} & 596\_fri\_ & breast\_ & \multirow{2}{*}{cars} & \multirow{2}{*}{irish} & 522\_ & \multirow{2}{*}{profb} & tic\_tac\_ & \multirow{2}{*}{churn} & \multirow{2}{*}{adult} & \\
         & & c2\_250 & cancer & & & pm10 & & toe & & & \\
        \hline
        DataSynthesizer-Ind & 0.37 & 0.05 & 0.34 & 0.34 & 0.39 & \cellcolor{gray!30}0.39 & 0.13 & 0.10 & 0.07 & \cellcolor{gray!30}0.40 & 2 \\
        DataSynthesizer-Cor & 0.36 & 0.07 & 0.35 & \cellcolor{gray!30}0.31 & 0.34 & \cellcolor{gray!30}0.35 & 0.12 & 0.10 & 0.07 & 0.52 & 2 \\
        Synthetic Data Vault-ML & 0.34 & \cellcolor{gray!30}0.04 & \cellcolor{gray!30}0.23 & 0.39 & \cellcolor{gray!30}0.25 & 0.47 & \cellcolor{gray!30}0.10 & \cellcolor{gray!30}0.09 & \cellcolor{gray!30}0.04 & 0.48 & 6 \\
        Synthetic Data Vault-GC & 0.30 & \cellcolor{gray!30}0.03 & \cellcolor{gray!30}0.23 & \cellcolor{gray!30}0.30 & \cellcolor{gray!30}0.25 & 0.42 & \cellcolor{gray!30}0.10 & \cellcolor{gray!30}0.09 & \cellcolor{gray!30}0.04 & 0.46 & 7 \\
        Synthetic Data Vault-GAN & \cellcolor{gray!30}0.26 & 0.05 & \cellcolor{gray!30}0.23 & \cellcolor{gray!30}0.29 & \cellcolor{gray!30}0.24 & 0.45 & \cellcolor{gray!30}0.11 & \cellcolor{gray!30}0.09 & \cellcolor{gray!30}0.04 & 0.46 & 7 \\ 
        Recursive Random Projection & 0.42 & \cellcolor{gray!30}0.04 & \cellcolor{gray!30}0.23 & 0.40 & 0.24 & 0.67 & \cellcolor{gray!30}0.10 & \cellcolor{gray!30}0.09 & 0.07 & 0.54 & 4 \\
        Howso Engine & \cellcolor{gray!30}0.27 & 0.05 & \cellcolor{gray!30}0.23 & 0.39 & \cellcolor{gray!30}0.24 & 0.54 & \cellcolor{gray!30}0.10 & \cellcolor{gray!30}0.09 & \cellcolor{gray!30}0.04 & 0.53 & 6 \\
    \end{tabular}
    \caption{Marginal distribution similarity score. The overall score is the mean of the Jensen-Shannon divergence scores calculated by equation~\ref{JSdivergence} through all features. The dark gray cell marks the algorithm in the highest rank by Scott-Knott analysis, and the light gray cell marks the algorithm in the second highest rank.}
    \label{tab:marginal}
\end{table*}

\subsection{Scott-Knott Analysis}
To perform significant test, we utilize the Scott-Knott Analysis. We choose Scott-Knott analysis since (a) it is fully non-parametric and (b) it can reduce the potential error during the analysis with only at most $O(log2(N))$ statistical tests for the $O(N^2)$ analysis.

In our experiment, we repeat each synthetic data generation algorithm 10 times since they are stochastic. With a list of candidates $C$, where each candidate $c_i$ is a list of results from 10 repeats for a certain synthetic data generation algorithm, Scott-Knott recursively partitions $C$ into two sub-lists $C_1$ and $C_2$. The split is based on the expected mean value before and after the division, which the goal is to maximize the expected mean value~\cite{xia2018hyperparameter, emblem, 9463120}. The delta of expected mean value before and after the split is calculated as follow:
\begin{equation}
    E(\Delta) = \frac{\text{l}(C_1) \cdot |\mu(C_1) - \mu(C)| + \text{l}(C_2) \cdot |\mu(C_2) - \mu(C)|}{\text{l(C)}}
\end{equation}
where $\text{l}$ is the length function to count the length of each list.

After the split is finished, Scott-Knott will then utilize the Cliff's Delta procedure to check if two sub-lists differ significantly by
\begin{equation}
    Delta = \frac{\sum\limits_{x \in C_1} \sum\limits_{y \in C_2} \left\{ \begin{array}{l}
                    +1, \mbox{   if $x > y$}\\
                    -1, \mbox{   if $x < y$}\\
                    0,  \mbox{   if $x = y$}
                \end{array} \right.}{|C_1||C_2|}
\end{equation}
More specifically, Cliff's Delta estimates the probability that a value in the sub-list $C_1$ is greater than a value in the sub-list $C_2$, and then minus the reverse probability~\cite{macbeth2011cliff}. If Cliff's Delta is equal or greater than 0.147 (i.e. $Delta \geq 0.147$) then it is not a small effect~\cite{hess2004robust}.

\section{Results}\label{results}
In this section, we present our experimental results, and answer RQs based on the results.

\textit{RQ1: When considering the privacy, which synthetic data generation algorithm can generate the synthetic data with the highest privacy preservation?} To evaluate the synthetic data in terms of privacy, we implement the \textit{Privacy Preservation} metric to evaluate the privacy score. As we described in Section~\ref{metric}, privacy preservation evaluates the distance from the synthetic data point to the density of the region of the $k$-nearest neighbors of the closest original data point. We calculate such distance ratio for all synthetic data points, and take the geometric mean as the final privacy preservation score for the synthetic dataset. As mentioned in Section~\ref{privacypreservation}, higher score refers to better privacy. However, the score of 1 already indicates that the synthetic data lies out of the density region of its closest original data point. Hence, in this research question, we consider an approach with good privacy preservation if its score is greater or equal than 1. Table~\ref{tab:anonymityPreservation} presents the experimental results. At first glance, DataSynthesizer has higher raw values than any other algorithms. However, those methods with score equal or greater than 1 will not induce the privacy issue in the real world application. Hence, if a method is not in the highest rank through statistical testing but its raw value is equal or greater than 1, we still consider it as a good method and marked them with grey color in Table~\ref{tab:anonymityPreservation}.

As we can see, synthetic data generated by DataSynthesizer has the highest privacy score. After that, Howso engine has promising performance on 6 case studies. SDV and recursive random projection do not perform well in terms of privacy preservation. Hence, our answer to RQ1 is

\textbf{DataSynthesizer and Howso engine are two promising algorithms that can generate synthetic data with good privacy preservation score. If we only consider privacy, the result would recommend DataSynthesizer. However, as we will discuss in the RQ5, the evaluation of synthetic data cannot only concentrate on the privacy, and therefore, we recommend Howso engine more than DataSynthesizer.}

\begin{table*}[t]
    \centering
    \begin{tabular}{c|c||c|c|c|c|c|c|c|c|c|c||c}
        \multirow{3}{*}{\textbf{Metric}} & \multirow{3}{*}{\textbf{Algorithm}} & \multicolumn{10}{c||}{\textbf{Benchmarks}} & \multirow{3}{*}{\textbf{Wins}} \\
        \cline{3-12}
         & & \multirow{2}{*}{glass} & 596\_fri\_ & breast\_ & \multirow{2}{*}{cars} & \multirow{2}{*}{irish} & 522\_ & \multirow{2}{*}{profb} & tic\_tac\_ & \multirow{2}{*}{churn} & \multirow{2}{*}{adult} & \\
         & & & c2\_250 & cancer & & & pm10 & & toe & & & \\
        \hline
        \hline
        \multirow{7}{*}{Accuracy} & DS\_Ind & 0.15 & - & 0.31 & 0.44 & 0.57 & - & 0.33 & 0.35 & 0.80 & 0.65 & 0 \\
         & DS\_Cor & 0.32 & - & 0.64 & 0.20 & 0.66 & - & 0.53 & 0.66 & 0.70 & 0.75 & 0 \\
         & SDV\_ML & 0.34 & - & 0.67 & 0.70 & 0.79 & - & \cellcolor{gray!30}0.67 & 0.63 & 0.85 & 0.76 & 1 \\
         & SDV\_GC & 0.41 & - & \cellcolor{gray!30}0.71 & 0.65 & 0.77 & - & \cellcolor{gray!30}0.67 & 0.65 & 0.85 & 0.74 & 2 \\
         & SDV\_GAN & 0.17 & - & 0.64 & 0.38 & 0.52 & - & 0.64 & 0.58 & 0.82 & 0.79 & 0 \\ 
         & RRP & \cellcolor{gray!30}0.71 & - & \cellcolor{gray!30}0.72 & \cellcolor{gray!30}0.84 & \cellcolor{gray!30}0.96 & - & 0.64 & 0.77 & 0.87 & \cellcolor{gray!30}0.82 & 5 \\
         & Howso Engine & \cellcolor{gray!30}0.76 & - & \cellcolor{gray!30}0.74 & \cellcolor{gray!30}0.86 & \cellcolor{gray!30}0.94 & - & \cellcolor{gray!30}0.67 & \cellcolor{gray!30}0.85 & \cellcolor{gray!30}0.90 & \cellcolor{gray!30}0.81 & 8 \\
         \hline
         \hline
        \multirow{7}{*}{Precision} & DS\_Ind & 0.02 & - & 0.10 & 0.58 & 0.64 & - & 0.11 & 0.47 & 0.74 & 0.61 & 0 \\
         & DS\_Cor & 0.23 & - & 0.60 & 0.45 & 0.70 & - & 0.47 & 0.61 & 0.73 & 0.69 & 0 \\
         & SDV\_ML & 0.26 & - & 0.50 & 0.68 & 0.79 & - & 0.60 & 0.52 & 0.74 & 0.70 & 0 \\
         & SDV\_GC & 0.44 & - & \cellcolor{gray!30}0.69 & 0.64 & 0.77 & - & 0.60 & 0.62 & 0.78 & 0.67 & 1 \\
         & SDV\_GAN & 0.22 & - & 0.58 & 0.42 & 0.56 & - & 0.60 & 0.58 & 0.81 & 0.79 & 0 \\ 
         & RRP & \cellcolor{gray!30}0.67 & - & \cellcolor{gray!30}0.65 & \cellcolor{gray!30}0.83 & \cellcolor{gray!30}0.96 & - & 0.61 & 0.78 & 0.87 & \cellcolor{gray!30}0.82 & 5 \\
         & Howso Engine & \cellcolor{gray!30}0.73 & - & \cellcolor{gray!30}0.73 & \cellcolor{gray!30}0.87 & \cellcolor{gray!30}0.94 & - & \cellcolor{gray!30}0.65 & \cellcolor{gray!30}0.86 & \cellcolor{gray!30}0.90 & 0.80 & 7 \\
         \hline
         \hline
        \multirow{7}{*}{Recall} & DS\_Ind & 0.15 & - & 0.31 & 0.44 & 0.57 & - & 0.33 & 0.35 & 0.80 & 0.65 & 0 \\
         & DS\_Cor & 0.17 & - & 0.64 & 0.20 & 0.66 & - & 0.53 & 0.66 & 0.70 & 0.75 & 0 \\
         & SDV\_ML & 0.34 & - & 0.67 & 0.70 & 0.79 & - & \cellcolor{gray!30}0.67 & 0.63 & 0.85 & 0.76 & 1 \\
         & SDV\_GC & 0.41 & - & \cellcolor{gray!30}0.71 & 0.65 & 0.77 & - & \cellcolor{gray!30}0.67 & 0.65 & 0.85 & 0.74 & 2 \\
         & SDV\_GAN & 0.17 & - & 0.64 & 0.38 & 0.52 & - & 0.64 & 0.58 & 0.82 & 0.79 & 0 \\ 
         & RRP & \cellcolor{gray!30}0.71 & - & \cellcolor{gray!30}0.72 & \cellcolor{gray!30}0.84 & \cellcolor{gray!30}0.96 & - & 0.64 & 0.77 & 0.87 & \cellcolor{gray!30}0.82 & 5 \\
         & Howso Engine & \cellcolor{gray!30}0.76 & - & \cellcolor{gray!30}0.74 & \cellcolor{gray!30}0.86 & \cellcolor{gray!30}0.94 & - & \cellcolor{gray!30}0.67 & \cellcolor{gray!30}0.85 & \cellcolor{gray!30}0.90 & \cellcolor{gray!30}0.81 & 8 \\
         \hline
         \hline
        \multirow{7}{*}{MCC} & DS\_Ind & 0.00 & - & 0.00 & 0.15 & 0.22 & - & 0.00 & -0.13 & -0.05 & -0.06 & 0 \\
         & DS\_Cor & 0.06 & - & 0.03 & -0.02 & 0.35 & - & -0.08 & 0.10 & -0.09 & 0.11 & 0 \\
         & SDV\_ML & -0.02 & - & 0.00 & 0.31 & 0.51 & - & 0.07 & -0.04 & 0.00 & 0.11 & 0 \\
         & SDV\_GC & 0.18 & - & 0.05 & 0.33 & 0.57 & - & \cellcolor{gray!30}0.10 & 0.15 & 0.03 & 0.06 & 1 \\
         & SDV\_GAN & 0.07 & - & 0.00 & -0.08 & 0.06 & - & 0.03 & 0.04 & 0.18 & 0.40 & 0 \\ 
         & RRP & \cellcolor{gray!30}0.60 & - & \cellcolor{gray!30}0.12 & \cellcolor{gray!30}0.65 & \cellcolor{gray!30}0.92 & - & \cellcolor{gray!30}0.13 & 0.44 & 0.31 & \cellcolor{gray!30}0.50 & 6 \\
         & Howso Engine & \cellcolor{gray!30}0.64 & - & \cellcolor{gray!30}0.34 & \cellcolor{gray!30}0.70 & 0.86 & - & \cellcolor{gray!30}0.17 & \cellcolor{gray!30}0.67 & \cellcolor{gray!30}0.50 & 0.45 & 6 \\
         \hline
         \hline
        \multirow{7}{*}{RMSE} & DS\_Ind & - & 1.23 & - & - & - & 1.42 & - & - & - & - & 0 \\
         & DS\_Cor & - & 1.13 & - & - & - & 1.05 & - & - & - & - & 0 \\
         & SDV\_ML & - & 0.91 & - & - & - & 0.84 & - & - & - & - & 0 \\
         & SDV\_GC & - & 0.87 & - & - & - & 2.35 & - & - & - & - & 0 \\
         & SDV\_GAN & - & 1.20 & - & - & - & 1.05 & - & - & - & - & 0 \\ 
         & RRP & - & 0.53 & - & - & - & \cellcolor{gray!30}0.73 & - & - & - & - & 1 \\
         & Howso Engine & - & \cellcolor{gray!30}0.46 & - & - & - & \cellcolor{gray!30}0.72 & - & - & - & - & 2 \\
         \hline
         \hline
        \multirow{7}{*}{R$^2$} & DS\_Ind & - & -0.59 & - & - & - & -1.51 & - & - & - & - & 0 \\
         & DS\_Cor & - & -0.33 & - & - & - & -0.39 & - & - & - & - & 0 \\
         & SDV\_ML & - & 0.18 & - & - & - & -0.03 & - & - & - & - & 0 \\
         & SDV\_GC & - & 0.14 & - & - & - & -6.88 & - & - & - & - & 0 \\
         & SDV\_GAN & - & -0.74 & - & - & - & -0.61 & - & - & - & - & 0 \\ 
         & RRP & - & 0.68 & - & - & - & \cellcolor{gray!30}0.31 & - & - & - & - & 1 \\
         & Howso Engine & - & \cellcolor{gray!30}0.77 & - & - & - & \cellcolor{gray!30}0.29 & - & - & - & - & 2 \\
         \hline
         \hline
        \multirow{7}{*}{Spearman} & DS\_Ind & - & 0.19 & - & - & - & 0.36 & - & - & - & - & 0 \\
         & DS\_Cor & - & 0.49 & - & - & - & 0.48 & - & - & - & - & 0 \\
         & SDV\_ML & - & 0.74 & - & - & - & 0.64 & - & - & - & - & 0 \\
         & SDV\_GC & - & 0.74 & - & - & - & 0.51 & - & - & - & - & 0 \\
         & SDV\_GAN & - & 0.48 & - & - & - & 0.47 & - & - & - & - & 0 \\ 
         & RRP & - & \cellcolor{gray!30}0.92 & - & - & - & \cellcolor{gray!30}0.77 & - & - & - & - & 2 \\
         & Howso Engine & - & \cellcolor{gray!30}0.94 & - & - & - & \cellcolor{gray!30}0.78 & - & - & - & - & 2 \\
    \end{tabular}
    \caption{Model comparison score. Classification tasks are evaluated by Accuracy, Precision, Recall, and MCC, while regression tasks are evaluated by RMSE, R$^2$, and Spearman. The dark gray cell marks the algorithm in the highest rank and the light gray cell marks the algorithm in the second highest rank. Due to space reason, DataSynthesizer is simplified to DS, Synthetic Data Vault is simplified to SDV, and Recursive Random Projection is simplified to RRP.}
    \label{tab:ModelComparison}
\end{table*}

\textit{RQ2: Which synthetic data generation algorithm can generate data that has higher similarity to the original data?} Similarity is another important measurement on checking if synthetic data can provide same information as original data does. We implement \textit{statistical similarity score} and \textit{marginal distribution similarity score} to evaluate the synthetic data. Specifically,
\begin{itemize}
    \item Statistical similarity score evaluate the synthetic data by comparing its statistical measurements to the original data for each feature.
    \item Marginal distribution similarity score evaluates the synthetic data from the estimated marginal probability distribution of each feature.
\end{itemize}

Table~\ref{tab:descriptiveStatistics} shows the statistical similarity score, which is calculated by evaluating the SMAPE of 14 different statistical measurements in synthetic data and original data (described in~\S\ref{statisticalsimilarity}). From this table, we can find Howso engine and recursive random projection algorithm have better performance in more case studies than any other algorithms. This indicates that the synthetic data generated by Howso engine and recursive random projection algorithm can better capture the statistical patterns in the original data. More specifically, the large SMAPE scores on DataSynthesizer reduce its advantage on the privacy preservation since the synthetic data generated by it does not follow the statistical patterns in the original data.

Table~\ref{tab:marginal} presents the marginal probability distribution similarity score, which is calculated by Jensen-Shannon divergence. Specifically, the marginal distribution is estimated through the KNN density estimation in numeric features and normalized value counts in categorical features. Different to the statistical similarity, this metric concentrates more on the estimated local probability distribution of each feature and check if the probability distribution of the same feature from synthetic data and orignial data is similar or not through Jenson-Shannon divergence. As we can see, most of the algorithms get good scores except DataSynthesizer, which is also shown to have worse score in statistical similarity score.

For all other algorithms except DataSynthesizer, we prefer Synthetic Data Vault with Gaussian Copula, recursive random projection, and Howso engine since they both perform well in two similarity metrics. Hence, our answer for RQ3 is

\textbf{Considering two different similarity measurements, Synthetic Data Vault with Gaussian Copula, recursive random projection, and Howso engine achieve good performance in both two metrics. Thus, we recommend these three algorithms when evaluating the similarity.}



\begin{figure}[!b]
    \centering
    \includegraphics[width=0.5\textwidth]{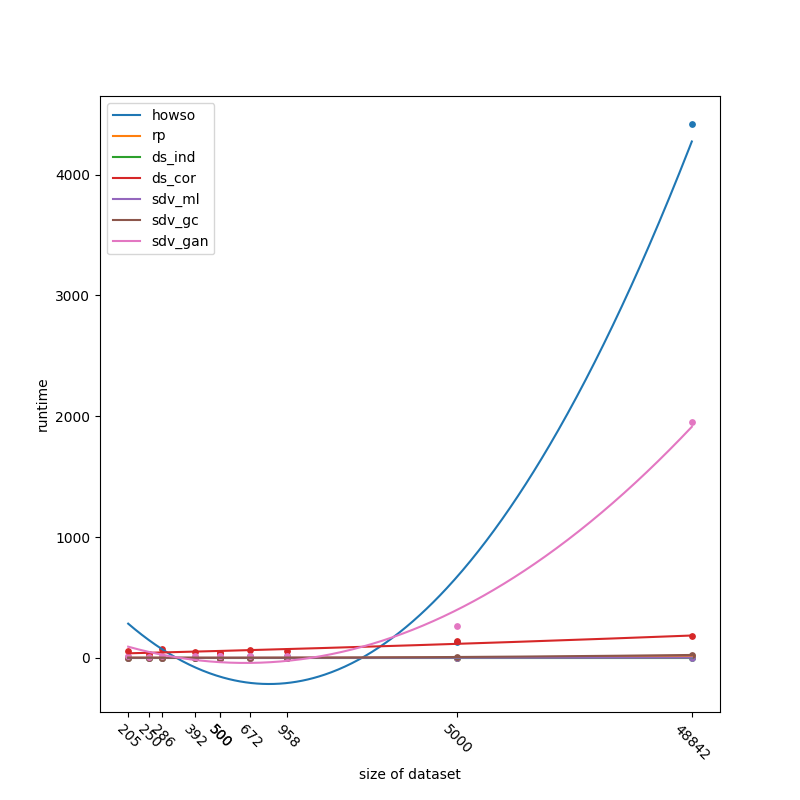}
    \caption{Runtime of different synthetic data approaches.}
    \label{fig:runtime}
\end{figure}

\textit{RQ3: When the machine learning model is trained on the synthetic data, can the model achieve compatible performance with those trained on the original data?} The model performance score is a very important indicator to evaluate the synthetic data. We split both the original data and the synthetic data to 80\% training set and 20\% test set. Then we use the model trained on the synthetic training data to predict the test set split from the original data. As stated in Section~\ref{metric}, for the classification task we collect accuracy, precision, recall, and Matthews correlation coefficient, and for the regression task we collect RMSE, R$^2$, and Spearman correlation coefficient. Table~\ref{tab:ModelComparison} shows the scores of these metrics. As we can see, in all metrics, Howso engine and recursive random projection framework get significant higher performance than other two state-of-the-art algorithms. Therefore, we conclude that the synthetic data generated by these two algorithms can be applied to the real world machine learning models without loss on the information. Based on above, our answer to RQ3 is:

\textbf{Howso engine and recursive random projection algorithm can generate synthetic data that does not loss any original information when training on the machine learning models.}

\textit{RQ4: Which algorithm has the best scalability?} To answer this question, we record the runtime of each \textbf{generation algorithm}. Figure~\ref{fig:runtime} shows the runtime for each generation algorithm. To visualize the scalability, we first record the actual runtime for each algorithm in each case study. The scatters shows the point $(x, y)$ where $x$ is the size of the dataset, and $y$ is the runtime. After that, we plot the best fitting polynomial curve, which presents the trending of the runtime when the size of the dataset is increasing. As we can see, when the size of the dataset is small, all algorithms have similar runtime. However, when the size of the dataset becomes larger (e.g. 40k+ rows), the runtime of Howso engine and GAN based algorithm increases exponentially (i.e. the blue line and the pink line). To the contrary, our proposed recursive random projection algorithm, along with DataSynthesizer and Synthetic Data Vault with Gaussian Copula, have better efficiency, which the best fitting curve of runtime is close to linear with low slope even though the size of the dataset goes exponentially large. Hence, our answer to RQ4 is:

\textbf{Our proposed recursive random projection framework, along with DataSynthesizer and Synthetic Data Vault with Gaussian Copula have the best scalability even when the size of the dataset is very large.}

\textit{RQ5: What recommendation can we provide from analyzing the conclusions from RQ1 to RQ4?}
Evaluating synthetic data needs to consider multi-dimensional criteria. To empirically evaluate all performance from RQ1 to RQ4, we transfer each criteria to a 0-1 range and use the radar chart to visualize the performance. Specifically
\begin{itemize}
    \item For \textit{privacy preservation}, all scores greater than 1 are treated as 1, and all other values will not be modified. The overall score of an algorithm will be the geometric mean of its scores in 10 case studies (If an algorithm has score 0, then we use 0.001 when calculating the geometric mean).
    \item For \textit{statistical similarity} and \textit{marginal distribution similarity}, the scores are 1 minus the values in Table~\ref{tab:descriptiveStatistics} since smaller is better in these two metrics. The overall score is also the geometric mean through 10 case studies.
    \item For \textit{model comparison}, the negative correlation coefficient will be treated as 0.001 since negative value means no correlation. We first calculate the geometric mean of each algorithm for each individual metric, and then calculate the geometric mean again for 7 metrics.
    \item For \textit{scalability}, we transfer the actual runtime by minmax scalar in each case study. Then the overall score will also be the geometric mean through 10 case studies for each individual algorithm. 
\end{itemize}

We select 4 algorithms to be presented in our chart. The first one is the correlation mode of DataSynthesizer. We do not choose independent mode since it is worse than correlation mode in privacy preservation, and similar to correlation mode in other metrics. The same procedure is applied to Synthetic Data Vault and the Gaussian Copula mode is selected. Hence in the chart, we have above two algorithms plus recursive random projection and Howso engine.

\begin{figure}
    \centering
    \includegraphics[width=0.5\textwidth]{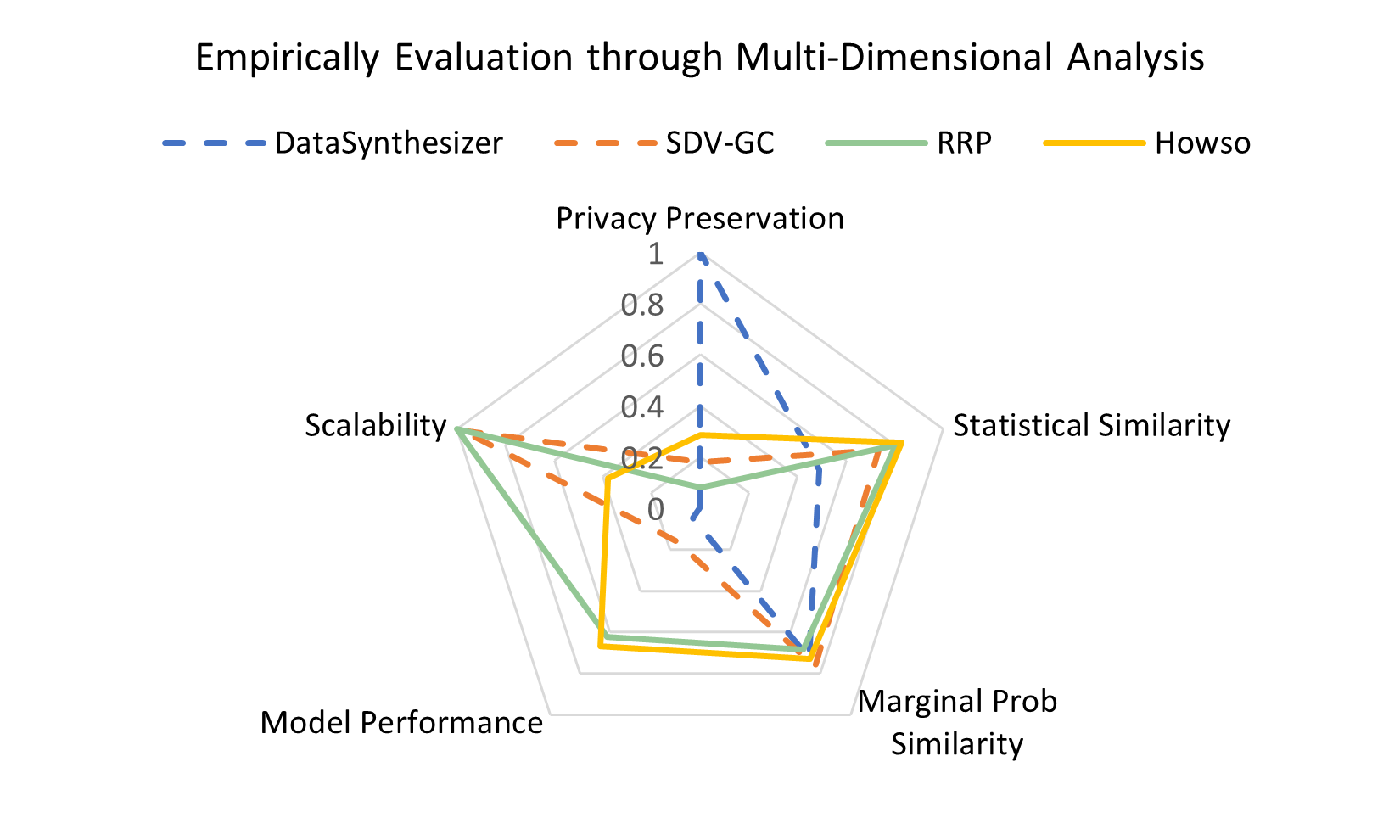}
    \caption{Radar chart of scores in multi-dimensional criteria.}
    \label{fig:radarchart}
\end{figure}

Figure~\ref{fig:radarchart} presents the radar chart of (a) DataSynthesizer with correlated mode, (b) SDV with Gaussain Copula, (c) Recursive Random Projection, and (d) Howso Engine. As we can see, the area covered by DataSynthesizer and SDV (blue dash line and orange dash line correspondingly) are obviously less than the area covered by Recursive Random Projection and Howso Engine. Hence, we offer two recommendations based on the analysis, and answer RQ5:
\begin{itemize}
    \item \textbf{As seen in Figure~\ref{fig:runtime}, runtime between different methods does not differ too much. Hence, if the scalability is not an issue, then we recommend Howso Engine since it can achieve higher accuracy score and promising privacy preservation score.}
    \item \textbf{However, if the dataset that needs to be synthesized is very large, and scalability becomes more important, we recommend our proposed Recursive Random Projection since it scales very fast and generate high accurate synthetic data.}
\end{itemize}

\section{Threat to Validity}\label{threat}
\textbf{Construct validity} mainly related to the different parameters setting and model construction which causes the different outcome. In our study, the threat of construct validity can happen in (a) the parameter choice in different generation models and (b) the choice of settings when evaluate the synthetic data. For example, the cluster size in our proposed random projection based framework can influence the performance of generating synthetic data. We empirically evaluate the different sizes of cluster and choose 12 as the parameter in our experiment. For another example, when we evaluate the synthetic data by training the machine learning model on it, we use the 80\% train test split on both original data and synthetic data. Different train test split ratio may cause different final outcome. The train test split ratio we used is the default setting which highly used in other machine learning studies. To reduce the threat, we build our experimental scripts as a python package, and allow researchers to replicate our experiment with their own parameter choice.

\textbf{Conclusion validity} refers to the threat that caused by applying different evaluation metrics when make the conclusion. To mitigate this threat,  we apply four metrics (privacy preservation, descriptive statistics, marginal probability, and model comparison) which include most of the evaluation aspects of synthetic data in the past literature. Researchers may expect different conclusion when applying different metrics on our methods.

\textbf{Internal validity} focuses on the correctness of treatment caused the outcome. To reduce the effect caused by this threat, we collect ten highly used machine learning benchmarks from PMLB and run all algorithms on those ten benchmarks. Also, we control the size of synthetic data and make it equal to the size of the original data.

\textbf{External validity} indicates the threat of applying this experiment to other fields. To mitigate this threat, the experimental goal of our study focuses on the machine learning object, which is one of the most well-known regions in the real world application. Moreover, our replication package can be applied to different datasets, which allows researchers to explore other real world application with our scripts.

\section{Conclusion}\label{conclusion}
In this study, we explore the synthetic data generation algorithms and discuss the different validation metrics. We proposed recursive random projection based generator, and compare it to (a) two state-of-the-art generation algorithms DataSynthesizer and Synthetic Data Vault, and (b) the Howso Engine from our industiral partner. We evaluate those synthetic data generators from (a) privacy preservation, (b) statistical similarity and marginal probability distribution similarity, (c) model performance comparison, and (d) scalability.

In privacy measurement, we find DataSynthesizer has the highest privacy preservation score through all case studies. Howso engine ranks behind the DataSynthesizer. However, when considering the similarity measurements and model comparison score, Howso engine and recursive random projection based framework get far more higher score than DataSynthesizer. Hence, we conclude that DataSynthesizer adds too much noise to the synthetic data which the fake points lie out of the original distribution and patterns.

By conducting the empirically analysis to five evaluation criteria, from the radar chart in RQ5, we offer two recommendations:
\begin{itemize}
    \item If scalability is not an issue, then we recommend Howso Engine which has highest accuracy performance and promising privacy score.
    \item However, when the dataset is large enough, which will cause the scalability issue, we recommend recursive random projection based framework since it scales fast and can achieve highest accuracy performance.
\end{itemize}

In the future work, we will explore more synthetic data generation algorithms, as well as more benchmarks. Moreover, current recursive random projection based framework does not particularly add differential privacy operators. In the future, we will design such operators based on the condition in each cluster, and improve its privacy preservation score.

\section*{Acknowledgment}
In this work, Howso funded NCState to comparatively assess numerous data synthesis methods. We assert  that the conclusions made here are the product of NCState and were not altered by our Howso collaborators.  

\bibliographystyle{IEEEtran}
\bibliography{ref}

\begin{IEEEbiography}[{\includegraphics[width=1in,height=1.25in,clip,keepaspectratio]{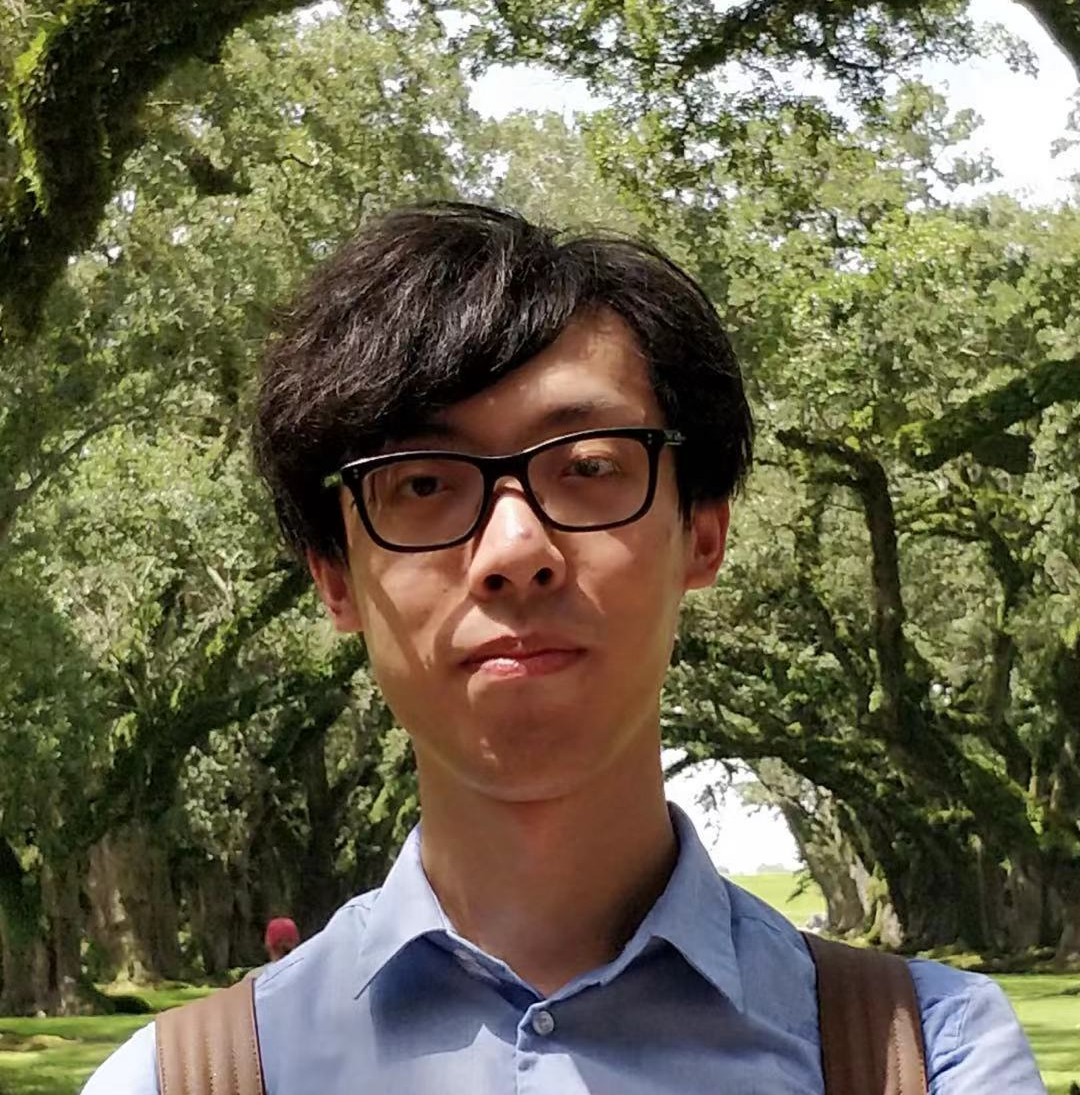}}]{Xiao Ling} is a fifth year PhD student in Computer Science at NC State University. His research interests include automated software testing, machine learning for software engineering, and landscape analysis for software analytics.
\end{IEEEbiography}

\begin{IEEEbiography}[{\includegraphics[width=1in,height=1.25in,clip,keepaspectratio]{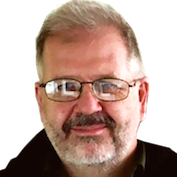}}]{Tim Menzies} (IEEE Fellow, Ph.D. UNSW, 1995)
is a Professor in computer science  at NC State University, USA,  
where he teaches software engineering,
automated software engineering,
and programming languages.
His research interests include software engineering (SE), data mining, artificial intelligence, and search-based SE, open access science. 
For more information,  please visit \url{http://timm.fyi}.
\end{IEEEbiography}

\begin{IEEEbiography}[{\includegraphics[width=1in,height=1.25in,clip,keepaspectratio]{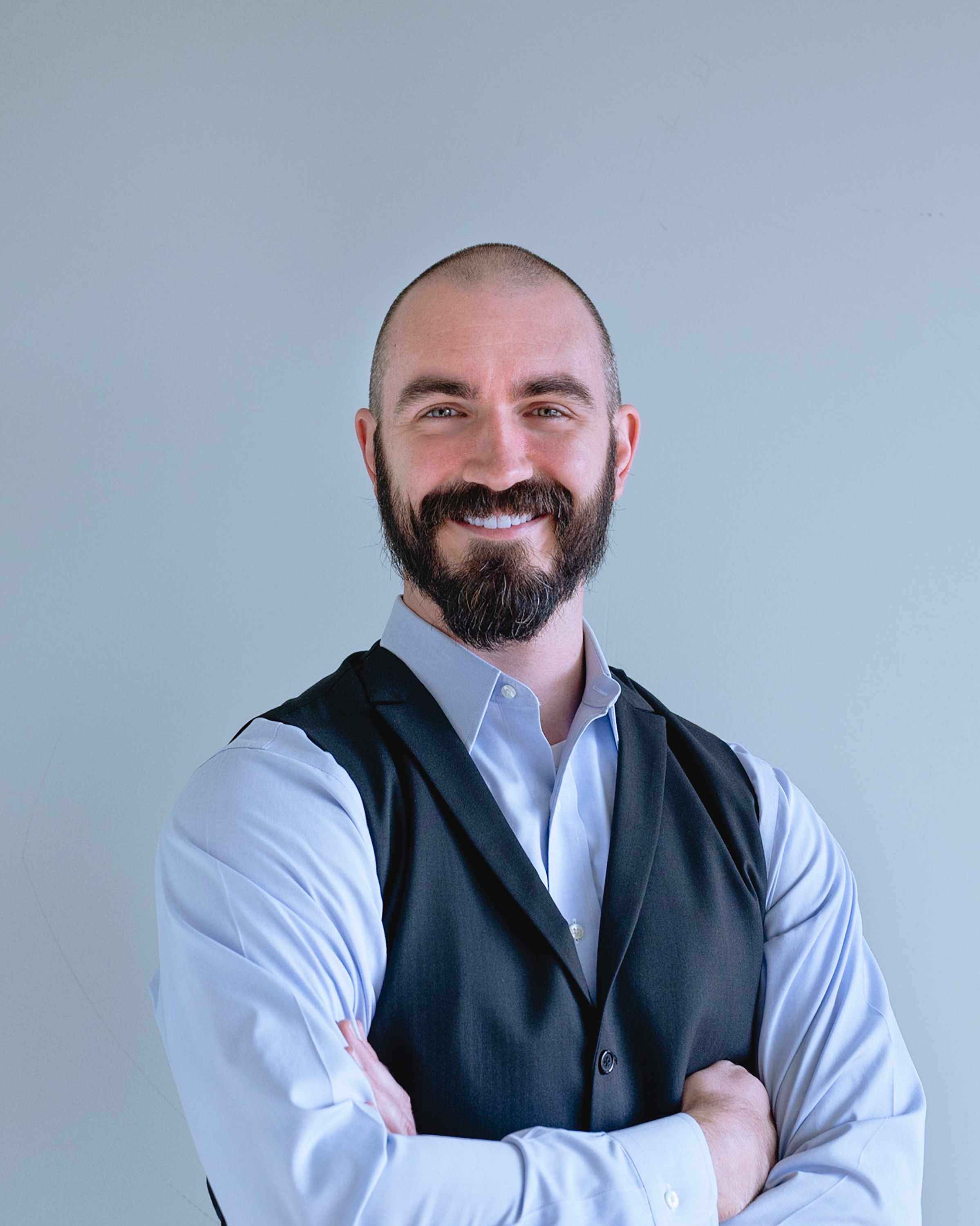}}]{Chris Hazard} is cofounder and CTO of the understandable and privacy enhancing AI company Howso.  Chris holds a PhD in computer science from NC State and has a long career across software, AI, and gaming including affiliations with Motorola, Hazardous Software, Kiva Systems (now Amazon Robotics), and NATO.
\end{IEEEbiography}

\begin{IEEEbiography}
[{\includegraphics[width=1in,height=1.25in,clip,keepaspectratio]{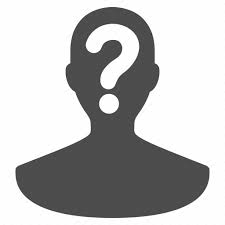}}]{Jack Shu} is the director of sales engineer of the understandable and privacy enhancing AI company Howso. Jack earns a Master's degree in Johns Hopkins Whiting School of Engineering and Bachelor's degree in University of Washington.
\end{IEEEbiography}

\begin{IEEEbiography}[{\includegraphics[width=1in,height=1.25in,clip,keepaspectratio]{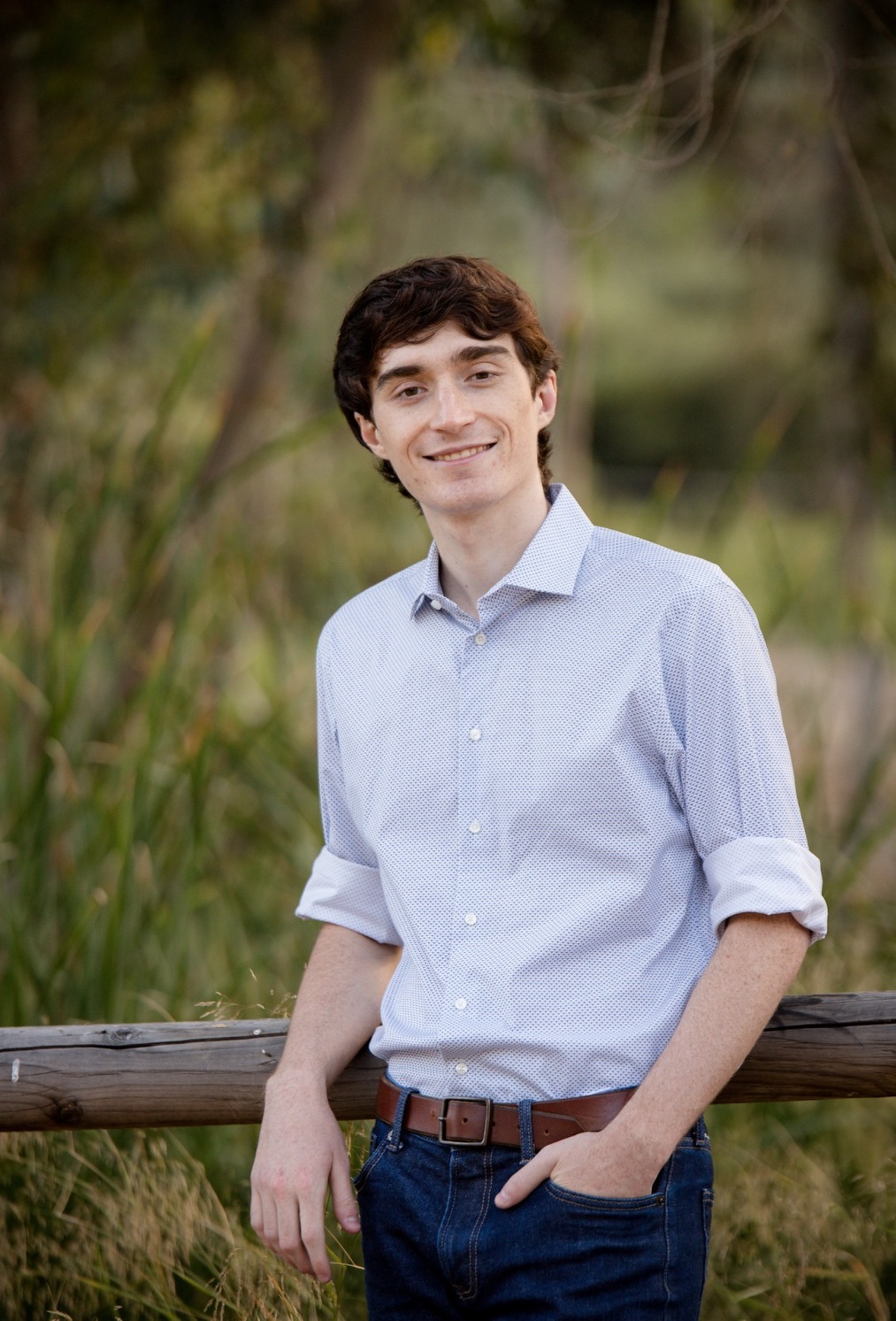}}]{Jacob Beel} attended the University of California, Irvine and then Georgia Tech earning a BS and MS in Computer Science, respectively. Deeply concerned about the social and ethical implications of AI, Jacob has been with Howso (formerly Diveplane) since graduating. There, he aims to bring interpretability and attributability to AI and discover novel ways of accomplishing a variety of tasks.

\end{IEEEbiography}

\EOD

\end{document}